\documentclass[preprintnumbers,amsmath,amssymb,usenatbib,nofootinbib,superscriptaddress,showkeys,showpacs,11pt]{revtex4-2}

\usepackage{amsmath}
\usepackage{amsfonts,color}
\usepackage{amssymb,float}
\usepackage{graphicx}
\usepackage{appendix}
\usepackage[colorlinks]{hyperref}
\usepackage{xcolor}
\usepackage{orcidlink}
\usepackage{epsf}
\usepackage{bm}
\usepackage{epstopdf}
\usepackage{natbib}
\usepackage{lipsum}
\usepackage{multirow}
\usepackage{nicematrix}

\begin{document}

\title{Correspondence between new agegraphic dark energy and Bose–Einstein condensate dark matter in the context of \(f(T)\) gravity}
\thanks{\bf Dedicated to Prof. Jafar Sadeghi in honor of his 60th jubilee}

\author{Alireza Amani\orcidlink{0000-0002-1296-614X}}
\email{al.amani@iau.ac.ir}
\affiliation{Department of Physics, Am.C., Islamic Azad University, Amol, Iran}

\date{\today}

\begin{abstract}
In this paper, we investigate the cosmic evolution within the framework of $f(T)$ gravity using a flat-FRW background and model the universe as consisting of three components: baryonic matter, dark matter, and dark energy. We consider the new agegraphic (NA) as an alternative for dark energy and the Bose-Einstein condensation (BEC) as an alternative for dark matter. After that, first we obtain the Friedman equations and then we obtain the continuity equations in the presence of the interaction term between the dark components of the universe, where the interaction term represents the energy flow from dark matter to dark energy. In what follows, we plot the variation of the cosmological parameters of dark energy in terms of the redshift parameter by using the power-law cosmology. Finally, we investigate the evolution and stability of the universe and report the values of the density parameters of the universe components which confirm the present model with observational data.

\end{abstract}


\maketitle
\newpage
%##########################################################
%##########################################################
\section{Introduction}\label{I}

Around the beginning of the new century, Riess et al \cite{Riess_1998} found out that the present universe is in an accelerated phase using type-Ia supernovae data, and then other groups \cite{Perlmutter_1999, Bennett_2003, Tegmark_2004} reconfirmed the same conclusion in the context of the cosmic microwave background radiation and the large-scale structure of the universe. The origin of these studies led to the definition of a mysterious energy called dark energy, which is responsible for the expansion of the universe. For this purpose, many models were studied in the fundamental theoretical framework such as string theory and quantum gravity, some of which can be mentioned such as the cosmological constant, modified gravity and teleparallel gravity \cite{Weinberg-1989, Ng-1992, Chiba-2000, Kamenshchik-2001, Caldwell-2002, Singh-2003,Guo-2005, Wei-2005, Setare1-2009, Amani-2011, Sen-2002, Bagla-2003, Sadeghi1-2009, Setare-2009, Amani-2013, Amani-2014, Battye-2016, Li-2012, Dolgov-2003, Faraoni-2006, Nojiri-2007, Iorio-2016, Li-2004, Wei-2009, Amani1-2011, Campo-2011, Hu-2015, Wei-2007, Jamil-2010, Jawad-2013, Shtanov-2003, Sadeghi-2009, Sadeghi-2010, Amani-2016, Singh-2016, Capozziello-2011, Myrzakulov-2011, Pourbagher_2019, Rezaei-2017, pourbagher1-2020, Sahni-2003, Setare-2008, Brito-2015}.

In recent years, modified theories of gravity have received widespread attention as promising frameworks for explaining the acceleration of the universe in recent times, without the need for dark energy of unknown nature. One of these models is the teleparallel gravity model, which was first proposed by Einstein to unify electromagnetism and gravity. In this theory, space-time is characterized by a linear connection without curvature and with twist by a metric tensor field called the dynamical tetrad field. Therefore, the geometry of general relativity, which is based on the Levi-Civita connection for a torsion-free curvature connection, gives way to the geometry of the teleparallel theory, which is based on the linear Weitzenb\"{o}ck connection in a curvature-free torsion connection. \cite{Einstein_1928, Weitzenbock_1923, Linder_2010, Myrzakulov_2011, Li_2011, Myrzakulov_2012, Harko_2011, Nojiri_2005}. Among these theories, $f(T)$ gravity, a generalization of teleparallel gravity in which the Lagrangian is defined as an arbitrary function of the torsion scalar $T$, offers a geometric structure and second-order field equations that distinguish it from higher-order theories such as $f(R)$. This torsion-based framework enables new cosmic dynamics and has been used in inflationary models, dark energy scenarios, and cosmic acceleration \cite{Capozziello-2022, Bahamonde-2023}.

On the other hand, the nature of dark matter remains one of the great mysteries of cosmology. While the standard cold dark matter (CDM) model has performed well on large scales, it faces challenges on galactic and subgalactic scales, such as the core-tip problem and the lack of satellites. An attractive alternative is the Bose-Einstein condensate (BEC) dark matter model, which models dark matter as a quantum fluid of ultralight bosons that condense on cosmic scales. This approach naturally introduces quantum pressure and internal potential into the model and can help smooth halo profiles and resolve small anomalies. However, dark matter is not visible in any spectrum of electromagnetic radiation, but it is an important topic in cosmology. Gravitational lensing and the cosmic microwave background are observational evidence of dark matter, as inferred from discrepancies between visible mass and gravitational effects observed in galaxy clusters. \cite{Kaiser-1993, Massey-2010, Seljak-1999, Arkani-2009}. Therefore, this experimental evidence shows that dark matter causes the gravitational effect between the formation and evolution of galaxies and clusters and large-scale structures. For this purpose, we use the Bose-Einstein condensation model to understand the concept of dark matter. Bose-Einstein condensation occurs in boson-diluted gases, where a large number of bosons simultaneously occupy a ground quantum state in a system. A number of papers have addressed the connection between BEC and dark matter. Therefore, we consider the BEC model as a viable candidate for describing dark matter \cite{Harko1-2011, Das-2015, Fukuyama-2008, Li-2014, Boehmer-2007, Das-2018, Suarez-2014, Chavanis-2012, Dev-2017, Velten-2012, Madarassy-2015, Kain-2010, Harko-2012, Harko-2019, Bettoni-2014, Zhang-2018, Harko-2015, Chavanis-2017, Harko1-2015, Chavanis-2017, Castellanos-2020, HajiSadeghi-2019, Atazadeh-2016, Mahichi-2021}.

The NA dark energy model is based on the quantum uncertainty principle and the energy-time relation in quantum mechanics. In this model, the dark energy density is related to the inverse square of cosmic time, but unlike the original version, conformal time is used instead of cosmic time. This modification makes the model more consistent with observational data and the dynamical behavior of the universe. We will describe the NA in more detail in Section \ref{IV} \cite{Karolyhazy-1966, Karolyhazy-1982, Karolyhazy-1986, Maziashvili-2007, Maziashvili1-2007}.

The combination of the $f(T)$ theory of gravity, Bose-Einstein condensed dark matter (BEC), and the NA dark energy model provides a new perspective for describing the universe. In the $f(T)$ theory, the curvature of space-time is replaced by torsion, which allows for geometric modifications to the field equations. On the other hand, BEC dark matter, with its quantum pressure and wave behavior, can shapes cosmic structures differently from classical cold dark matter. The NA model also describes dark energy using the time scale of the universe and the quantum uncertainty principle, which is naturally consistent with the properties of the BEC.

By combining these three components, it is possible to build a model in which the geometric torsion $f(T)$ affects the behavior of quantum dark matter, while dark energy appears dynamically via the agegraphic time scale. This synergy not only allows for a description of the cosmic acceleration without the need for a cosmological constant, but can also have observable effects on gravitational lensing data, large-scale structure, and fluctuations in the cosmic microwave background.

The main advantage of this approach is in creating a coherent, multi-layered cosmological framework that takes into account both the geometric aspects (torsion), the quantum properties of dark matter, and the dynamics of dark energy, and this could lead to a deeper understanding of the nature of the dark matter components of the Universe.

In an effort to comprehensively understand the nature of the accelerating universe, this research is based on three complementary frameworks: NA dark energy, BEC dark matter, and \(f(T)\) gravity. It is hypothesized that the interaction between these three components can provide a coherent and multi-layered picture of the evolution of the universe, such that the geometric torsion in \(f(T)\)gravity affects the dynamical behavior of BEC-type dark matter, while dark energy emerges in a quantum dynamical form and on a homomorphic time scale in the NA model. This theoretical equivalence not only allows for the description of cosmic acceleration without the need for the cosmological constant $\Lambda$, but also can yield results consistent with observational data such as the Hubble parameter, the cosmic microwave background radiation, and large-scale structures. Accordingly, the aim of this research is to reconstruct and analyze the behavior of the universe in the framework of \(f(T)\) by considering BEC as dark matter and NA as dark energy. For this purpose, the Friedmann and continuity equations are extracted for each component and the effect of the interaction term \(Q\) on the energy exchange between the dark sectors is investigated. Also, by applying power law cosmology and reconstructing the function \(f(T)\) in terms of the torsion scalar from observational data, thermodynamic stability and the transition from the matter-dominated to the acceleration-dominated phase are investigated. Finally, the agreement of the obtained values of the density parameters $\Omega_{bm_0}$, $\Omega_{dm_0}$ and $\Omega_{de_0}$ with the observed values is a confirmation of the efficiency of the proposed model in describing the current universe as a dynamic, geometric and quantum system.

This article is set up as follows:

In Sec. \ref{II}, we review the fundamental formulation of $f(T)$ gravity in the flat-FRW metric. In Sec. \ref{III}, we consider an interacting model between the dark parts of the universe. In Sec. \ref{IV}, we review the models of the NA dark energy and BEC dark matter, and we respectively obtain  the energy density of the NA dark energy and Equation of State (EoS) for BEC dark matter. In Sec. \ref{V}, we reconstruct the model in terms of the redshift parameter and establish a correspondence between the dark energy and dark matter components. Finally, in Sec. \ref{VI}, we will present a brief explanation for the present model.

%$$$$$$$$$$$$$$$$$$$$$$$$$$$$$$$
%$$$$$$$$$$$$$$$$$$$$$$$$$$$$$$$

\section{Covariant $f(T)$ gravity background}\label{II}

In this section, we intend to review modified teleparallel gravity by $f(T)$ gravity background within the covariant formulation. This formulation is able to guarantee local Lorentz invariance and frame independence. If we consider the model of $f(T)$ gravity only with the traditional pure-tetrad approach without explicitly introducing a spin connection, it violates the local Lorentz symmetry. That is, the investigation of the theory depends on the specific choice of tetrad, which can lead to different results with different choices. To avoid this problem, we include both the tetrad $e^a_\mu$ and a non-trivial spin connection $\omega^\lambda_{a \nu}$ in the theory, in order to define a covariant derivative and construct the torsion tensor, which can lead to Lorentz invariance. Therefore, the resulting equations are consistent with the physical concept, and are free of artifacts related to the tetrad frame.

Now, we start from the theory action of $f(T)$ gravity as follows:
\begin{equation}\label{action1}
S = \int e \left(\frac{f(T)}{2 \kappa^2} + \mathcal{L}_m \right) d^4x,
\end{equation}
where $\kappa^2=8 \pi G$ is the Einstein gravitational constant, $e=det\left( e^i_{\,\,\mu}\right)$ is the determinant of the tetrad field $e^i_{\,\,\mu}$, and $\mathcal{L}_m$ is the matter Lagrangian density. In addition, we can transform metric tensor $g_{\mu \nu}$ into the tetrad field $e^i_{\,\,\mu}$ in the form of $g_{\mu \nu}=\eta_{\mu \nu} e^i_{\,\,\mu} e^i_{\,\,\nu}$ ($\eta_{ij}=diag(+1,-1,-1,-1)$ is Minkowski space-time), where the relations $e_i\,^\mu e^i\,_\nu=\delta_\nu^\mu$ and $e_i\,^\mu e^j\,_\mu=\delta_i^j$ hold. Note that Greek and Latin letters represent space-time components and  tangent space-time components, respectively. The torsion scalar $T$ is defined as
\begin{equation}\label{torsion1}
  T = {S_\lambda }^{\mu \nu }\,{T^\lambda }_{\mu \nu },
\end{equation}
where the torsion tensor ${T^\lambda }_{\mu \nu }$, the antisymmetric tensor $S ^{\mu \nu }{}_\lambda$, and the contortion tensor $K ^{\mu \nu }{}_\lambda$ are as follows:
\begin{subequations}\label{torsion2}
\begin{eqnarray}
&{T^\lambda }_{\mu \nu } = {e_a}^\lambda \,\left( {{\partial _\mu }{e^a}_\nu  - {\partial _\nu }{e^a}_\mu } + \omega^{a}{}_{b\mu} e^{b}{}_{\nu} - \omega^{a}{}_{b\nu} e^{b}{}_{\mu}\right), \label{torsion2-1}\\
&{S_\lambda} ^{\mu \nu } = \frac{1}{2}\,\left( {K ^{\mu \nu }{}_\lambda + \delta _\lambda ^\mu \,T ^{\alpha \nu }{}_\alpha - \delta _\lambda ^\nu \,T ^{\alpha \mu }{}_\alpha} \right), \label{torsion2-2}\\
 &K^{\mu\nu}{}_{\lambda} =  - \frac{1}{2}\,\left( {{T^{\mu \nu }}_\lambda  - {T ^{\nu \mu }}_\lambda - {T_\lambda} ^{\mu \nu }} \right), \label{torsion2-3}
 \end{eqnarray}
\end{subequations}

The variation of the action with respect to the tetrad yields the Einstein's field equations:
\begin{eqnarray}\label{Eq4}
e^{-1} \partial_{\mu}\left(e\, e^\lambda_a\, S_{\lambda}^{~\mu\nu}\right) f_{T} - e_{a}^{\rho} \, T^{\lambda}_{~\mu\rho} S_{\lambda}{}^{\nu\mu} f_{T} + e^\lambda_a S_{\lambda}{}^{\mu\nu}\, \partial_{\mu}(T) f_{TT} +\omega^{\lambda}_{~a \nu} S_{\lambda}{}^{\mu\nu} f_{T} +\frac{1}{4}e_{a}^{\nu} f = \frac{1}{2} \kappa^2 e_{a}^{\lambda} \mathcal{T}_{\lambda}{}^{\nu},
\end{eqnarray}
where the indices $T$ and $TT$ indicate the first and second order derivatives with respect to torsion scalar, and $\mathcal{T}_{\lambda}{}^{\nu}$ is the energy-momentum tensor.

In the covariant formulation, the spin connection $\omega^\lambda_{a \nu}$ is chosen such that the total curvature vanishes, i.e., $R^{\lambda}_{~b\mu\nu}(\omega)=0$. This condition ensures that the geometry remains teleparallel (For more details, refer to Ref. \cite{Krssak-2016}).

In next section, we explore an interacting model in context of $f(T)$ gravity in Friedmann-Robertson-Walker (FRW) metric.

%$$$$$$$$$$$$$$$$$$$$$$$$$$$$$$$$$$$$$$$$$$$$$$$$$$$$$$$$$$$$$$$$$$$$$$$$$$$$$$$$$$$$$$$$$$$$$$$$$$$$$$$$$$$$$$
%$$$$$$$$$$$$$$$$$$$$$$$$$$$$$$$$$$$$$$$$$$$$$$$$$$$$$$$$$$$$$$$$$$$$$$$$$$$$$$$$$$$$$$$$$$$$$$$$$$$$$$$$$$$$$$

\section{the modified Friedmann equations and the continuity equations}\label{III}

According to the previous section, we continue the corresponding system with regard to the homogeneity and isotropy of the universe. For this purpose, we take the background manifold as flat-FRW background as
\begin{equation}\label{Eq6}
 ds^{2} = dt^{2} - a^{2}(t) \left(dx^2 +dy^2 + dz^2 \right),
\end{equation}
where $a(t)$ is the scale factor. The diagonal tetrad field is obtained as $e_{\mu}^a=\mathrm{diag}(1,a,a,a)$ or $e^{\mu}_a=\mathrm{diag}(1,a^{-1},a^{-1},a^{-1})$ that is compatible with a vanishing spin connection, making it a "good tetrad" in the covariant sense. The torsion scalar is as
\begin{eqnarray}\label{Eq7}
T = -6 H^2,
\end{eqnarray}
where $H = \dot{a}/{a}$ is the Hubble parameter. The energy-momentum tensor is obtained as follows:
\begin{equation}\label{Eq8}
\mathcal{T}_i^j=(\rho_{tot} + p_{tot}) u_i u^j - p_{tot}\,  \delta_i^j,
\end{equation}
where $\rho_{tot}$ and $p_{tot}$ are respectively the total energy density and the total pressure of fluid inside the universe. Since 4-velocity is a 4-vector, one is parallel to the direction of the time, i.e., $u^i$ = (+1,0,0,0), so we will have $u_i u^j$ = 1. In that case, the non-zero elements of energy-momentum tensor are obtained in the following form
\begin{equation}\label{Eq9}
 \mathcal{T}_i^j =diag \left( \rho_{tot}, - p_{tot}, - p_{tot}, - p_{tot}\right).
\end{equation}

Now according to the above, we can obtain the Friedmann equations as
\begin{subequations}\label{Eq10}
\begin{eqnarray}
 & \rho_{tot} = \frac{6}{\kappa^2} H^2 \partial_T f + \frac{1}{2 \kappa^2} f,\label{fried1-1}\\
 & {p}_{tot} = -\frac{2}{\kappa^2} \left(\dot{H} + 3 H^2 \right) \partial_T f + \frac{24}{\kappa^2} H^2 \dot{H} \partial_{T T} f - \frac{1}{2 \kappa^2} f, \label{fried1-2}
\end{eqnarray}
\end{subequations}
where the dot represents the derivative with respect to time evolution. The interesting point is that for the special case of $f(T) = T$, the above Friedman equation is the same as the standard Friedman equation in general relativity.

In what follows, we consider the content of the universe to be dominated by dark matter, baryonic matter and dark energy, the energy density and pressure of the universe can be written as three components of dark matter, baryonic matter and dark energy as follows:
\begin{subequations}\label{Eq11}
\begin{eqnarray}
& \rho_{tot} = \rho_{bm} + \rho_{dm} + \rho_{de},\label{Eq11-1}\\
& {p}_{tot} = p_{bm} + p_{dm} + p_{de},\label{Eq11-2}
\end{eqnarray}
\end{subequations}
where indices $bm$, $dm$ and $de$ are introduced as the contents of baryonic matter, dark matter and dark energy, respectively. Hence, we can obtain the total continuity equation from Eqs. \eqref{Eq10} as
\begin{equation}\label{Eq12}
\dot{\rho}_{tot}+3 H \left(\rho_{tot} + {p}_{tot}\right)=0,
\end{equation}
where continuity equations for the universe components with the presence of an interaction term between only dark energy and dark matter are obtained as follows:
\begin{subequations}\label{Eq13}
\begin{eqnarray}
&\dot{\rho}_{bm}+3 H \left(\rho_{bm} + p_{bm} \right)= 0,\label{Eq13-1}\\
&\dot{\rho}_{dm}+3 H \left(\rho_{dm} + p_{dm} \right)= Q,\label{Eq13-2}\\
&\dot{\rho}_{de}+3 H \left(\rho_{de} + {p}_{de}\right)= -Q,\label{Eq13-3}
\end{eqnarray}
\end{subequations}
where $Q$ is the interaction term. The interaction term appears in the universe when there is a flow of energy between the components of the universe, so that the direction of the energy flow is from dark energy to dark matter and there is no interaction between baryonic matter and other components. With this approach, the exclusive interaction between dark matter and dark energy without the involvement of visible matter leads to fascinating cosmological consequences. Note that the interaction term is confirmed from the point of view of the second law of thermodynamics and its value is positive \cite{Pavon_2009}. From the continuity equation, we find that the unit of term $Q$ is equal to the product of the Hubble parameter in the energy density. Therefore, in this work, we take the corresponding interaction term as $Q = 3 b^2 H \rho_{dm}$ in which $b$ is intensity of the energy flow transfer between the corresponding components of the universe. It can be motivated by field-theoretic considerations, such as coupling between scalar fields or effective interactions in modified gravity. The presence of $Q$ allows for dynamical adjustment of energy densities, potentially resolving the cosmic coincidence problem and influencing the effective EoS.

From Eqs. \eqref{Eq13-1} and \eqref{Eq13-2} we find
\begin{equation}\label{Eq13-11}
{\rho}_{bm} = \rho_{bm_0} a^{-3(1+\omega_{bm})},
\end{equation}
where $\omega_{bm} = p_{bm}/\rho_{bm}$ is EoS of baryonic matter which is a constant, and $\rho_{bm_0}$ is the present dark energy of baryonic matter.

After that, we obtain dark energy contribution components from Eqs. \eqref{Eq10} and \eqref{Eq11} as
\begin{subequations}\label{Eq14}
\begin{eqnarray}
& \kappa^2 \rho_{de} = 6 H^2 \partial_T f + \frac{1}{2} f - \kappa^2 \rho_{bm}- \kappa^2 \rho_{dm},\label{Eq14-1}\\
& \kappa^2 {p}_{de} = -2 \left(\dot{H} + 3 H^2 \right) \partial_T f + 24 H^2 \dot{H} \partial_{T T} f - \frac{1}{2} f - \kappa^2 \omega_{bm} \rho_{bm} - \kappa^2 \omega_{dm} \rho_{dm}, \label{Eq14-2}
\end{eqnarray}
\end{subequations}
where $\omega_{dm} = p_{dm}/\rho_{dm}$ is as the matter EoS parameter whose value is constant. Also, we can write down the EoS of dark energy as follows:
\begin{equation}\label{Eq15}
\omega_{de} = \frac{p_{de}}{\rho_{de}} = -1 - \frac{2 \dot{H} \partial_T f - 24 H^2 \dot{H} \partial_{T T} f + \kappa^2 \rho_{bm} \, (1+ \omega_{bm}) + \kappa^2 \rho_{dm} \, (1+ \omega_{dm})}{6 H^2 \partial_T f + \frac{1}{2} f - \kappa^2 \rho_{bm} - \kappa^2 \rho_{dm}},
\end{equation}
where the obtained EoS of dark energy is a modified form, this means that if we choose $f(T)=T$ in the absence of matter, it becomes $\omega=\frac{p_{tot}}{\rho_{tot}}=-1-\frac{2 \dot{H}}{3 H^2}$, which is same as the standard form in general relativity. Then, the obtained EoS shows us that the nature of the universe depends on $f(T)$ gravity and matter.

%################################################
%$$$$$$$$$$$$$$$$$$$$$$$$$$$$$$$$$$$$$$$$$$$$$$$$
\section{NA dark energy and BEC dark matter models}\label{IV}

In this section, we will examine the two main dark components in the present cosmological model: NA dark energy and BEC dark matter. Both components are rooted in quantum mechanical principles and play a complementary role in describing cosmic acceleration in the modified gravity framework.

First, we give a general overview of the subject of the NA dark energy. As we know the basis of quantum mechanics is based on the fact that things that can be measured have a special meaning in physics. Therefore, experimental observations can be the result of thoughts related to the specific science. In that case, the measurements of physical quantities in general relativity and quantum mechanics are different from each other in the sense that in the former, quantities are measured without any restrictions, but in the latter, the measurements are restricted and occur in a discrete manner due to quantum effects. Nevertheless, the theory of gravity that incorporates the principles of quantum mechanics is known as the quantum theory of gravity. Although the theory of quantum gravity is not fully established, some concepts based on the principles of quantum gravity are presented. This means that when gravitational fields such as the inflation period and the nature of black holes and quantum effects in gravity are very strong, they can be investigated using quantum gravity theory. Therefore, the phenomenology of space-time quantum fluctuations, which is explored by the theory of quantum gravity, is introduced as an agegraphic model. Accordingly, using the concept of space-time quantum fluctuations, the distance $t$ (with the speed of light $c = 1$) in Minkowski space-time was determined according to the uncertainty principle with the following accuracy
\begin{equation}\label{ٍEq16}
\delta\,t=\lambda\,{t_{{p}}}^{2/3}{t^{1/3}},
\end{equation}
where  $t_{{p}}$  is Planck's reduced time and $\lambda$  is a dimensionless constant of order unity \cite{Karolyhazy-1966, Karolyhazy-1982, Karolyhazy-1986, Maziashvili-2007, Maziashvili1-2007}. The energy density obtained from the NA model is equal to $\rho_{{DE}} \approx {\frac {1}{{\kappa}^{2}{t}^{2}}}$, where $t$ is the age of the universe. By substituting the age of the universe $t$ in the energy density of the NA dark energy model with conformal time $\tau$, the energy density of the NA dark energy model is obtained as follows:
\begin{equation}\label{Eq17}
\tau  = \int {\frac{{dt}}{a} = \int_0^a {\frac{{da}}{H{a^2}}} },
\end{equation}
 where $\dot \tau  = \frac{1}{a}$. Therefore, the energy density of the NA dark energy reads
\begin{equation}\label{Eq18}
{\rho _{\Lambda}} = \,\frac{{3{n^2}}}{{{\kappa^2}{\tau ^2}}},
\end{equation}
where $n^2$ is constant. 

On the other hand, we study the nature of dark matter by the regime of BEC. BEC is a state of matter in which the dilute Bose gas is cooled to such a low temperature that a phase transition occurs at this point and the macroscopic quantum phenomenon occurs as a quantum minimum state. Since dark matter has a significant contribution to the evolution of the universe, it leads us to adopt the BEC approach instead of dark matter. This means that dark matter is a type of BEC boson gas that dominates a part of the universe. Therefore, we consider the number density of particles by the Bose-Einstein statistics, where these particles form through the decoupling of the remaining plasma in the early universe. However, the energy density of dark matter is equal to the mass of dark matter multiplied by the number density of particles, and dark matter pressure is defined by Bose-Einstein statistics in a sphere with a radius of particles momentum (refer to \cite{Hogan-2000, Madsen-2001, Harko-2011, Craciun-2020, Mahichi_2023} for more details). In that case, by the generalized Gross-Pitaevskii equation \cite{Pitaevskii_2003, Pethick_2008}, we write down the dark matter pressure in terms of the energy density of dark matter as
\begin{equation}\label{Eq19}
p_{dm}= \omega_{dm} \rho_{dm}^2,
\end{equation}
where $\omega_{dm} = 2 \pi l_a /m_{dm}^3$, in which $l_a$ is the s-wave scattering length which here is introduced as dark matter EoS. By inserting Eq. \eqref{Eq19} into Eq. \eqref{Eq13-2}, we have
\begin{equation}\label{Eq20}
\rho_{dm}= \frac{1-b^2}{ (1-b^2) a^{3(1-b^2)} - \omega_{dm}},
\end{equation}
where by inserting $a_0=1$ ($a_0$ is the current scale factor) we obtain the current $\rho_{dm_0}$ as
\begin{equation}\label{Eq21}
\rho_{dm_0} = \frac{1-b^2}{1-b^2 - \omega_{dm}},
\end{equation}
 in that case,  $\rho_{dm_0}$ reads
 \begin{equation}\label{Eq22}
\rho_{dm} = \rho_{dm_0}\, \frac{1-b^2-\omega_{dm}}{ (1-b^2) a^{3(1-b^2)} - \omega_{dm}},
\end{equation}
 where this relationship shows that the energy density of dark matter depends on the scale factor, interacting term, and BEC dark matter.

%$$$$$$$$$$$$$$$$$$$$$$$$$$$$$$$$$$$$$$$$$$$
%$$$$$$$$$$$$$$$$$$$$$$$$$$$$$$$$$$$$$$$$$$$
\section{Reconstruction and correspondence of the model}\label{V}

In this section, we study the present model based on reconstructing and correspondence methods. In that case, we reconstruct the Friedmann equation in terms of redshift parameter, and then we correspond the NA dark energy with $f(T)$ gravity. For this purpose, we write down the transformation between the scale factor and redshift parameter $z$ as $a(t) = a_0 / (1+z)$ in which $a_0=1$ is the present scale factor, also differential form is as
\begin{equation}\label{Eq23}
\frac{d}{dt} = - H (1+z) \frac{d}{dz}.
\end{equation}

Now, in order to motivate the reconstruction of the function $f(T)$, we provide the following scientific explanation to proceed. We note that the modified Friedmann equations in the context of $f(T)$ gravity contain several degrees of freedom, including the energy density and pressure of the distinct components, the Hubble parameter, and the arbitrary function $f(T)$. Without additional constraints, this system of differential equations admits an infinite number of solutions. Therefore, in this work, we use the scale factor $a(t)$, the BEC model for dark matter, and the NA dark energy to reconstruct the function $f(T)$ along with physical motivation and consistency with observational data. These components act as guiding principles to reduce the degeneracy of the solutions and allow us to express the function $f(T)$ explicitly in terms of the torsion scalar $T$. This reconstruction not only increases the predictive power of the model, but also provides deeper insight into the interplay between geometry and quantum matter in the evolution of the universe.

Now in order to solve the model we consider a specific model called the power-law universe in the following form
\begin{equation}\label{Eq24}
  a(t) = {a_0} {\left(\frac{t}{t_0}\right)^m},
\end{equation}
where $t_0$ is the present age of the universe, and $m$ is dimensionless positive coefficient. The choice of a power-law scale factor is motivated by both theoretical and observational considerations. Power-law cosmology has been widely used in inflationary models, scalar field dynamics, and modified gravity scenarios due to their analytical tractability and ability to capture key features of cosmic evolution. Observational studies have shown that power-law behavior can approximate the expansion history in certain redshift ranges. By adopting this form, we reduce the arbitrariness in selecting a functional form for $f(T)$ and instead reconstruct it based on physically motivated components and observational constraints. It is important to note that the best-fit procedure primarily constrains the expansion history via the scale factor. This constraint indirectly informs the reconstruction of $f(T)$, which is derived from the modified Friedmann equations and the continuity relations. Therefore, while the fit does not directly test the gravity model, it provides a consistent framework for extracting $f(T)$ from observationally motivated dynamics. In that case, the Hubble's parameter and consequently the age of the universe will be:
\begin{subequations}\label{Eq25}
 \begin{eqnarray}
 H = \frac{m}{t}, \\ \label{Eq25-1}
 t_0 = \frac{m}{H_0}, \label{Eq25-2}
\end{eqnarray}
\end{subequations}
where $H(t_0)=H_0=67.4 \pm 0.5 \, km\,s^{-1}\,Mpc^{-1}$ \cite{Aghanim-2017}. In that case, we can obtain the Hubble parameter in terms of $z$ as
\begin{equation}\label{Eq26}
   H(z)=H_0  (1+z)^{\frac{1}{m}}.
\end{equation}

\begin{figure}[h]
\begin{center}
\includegraphics[scale=.4]{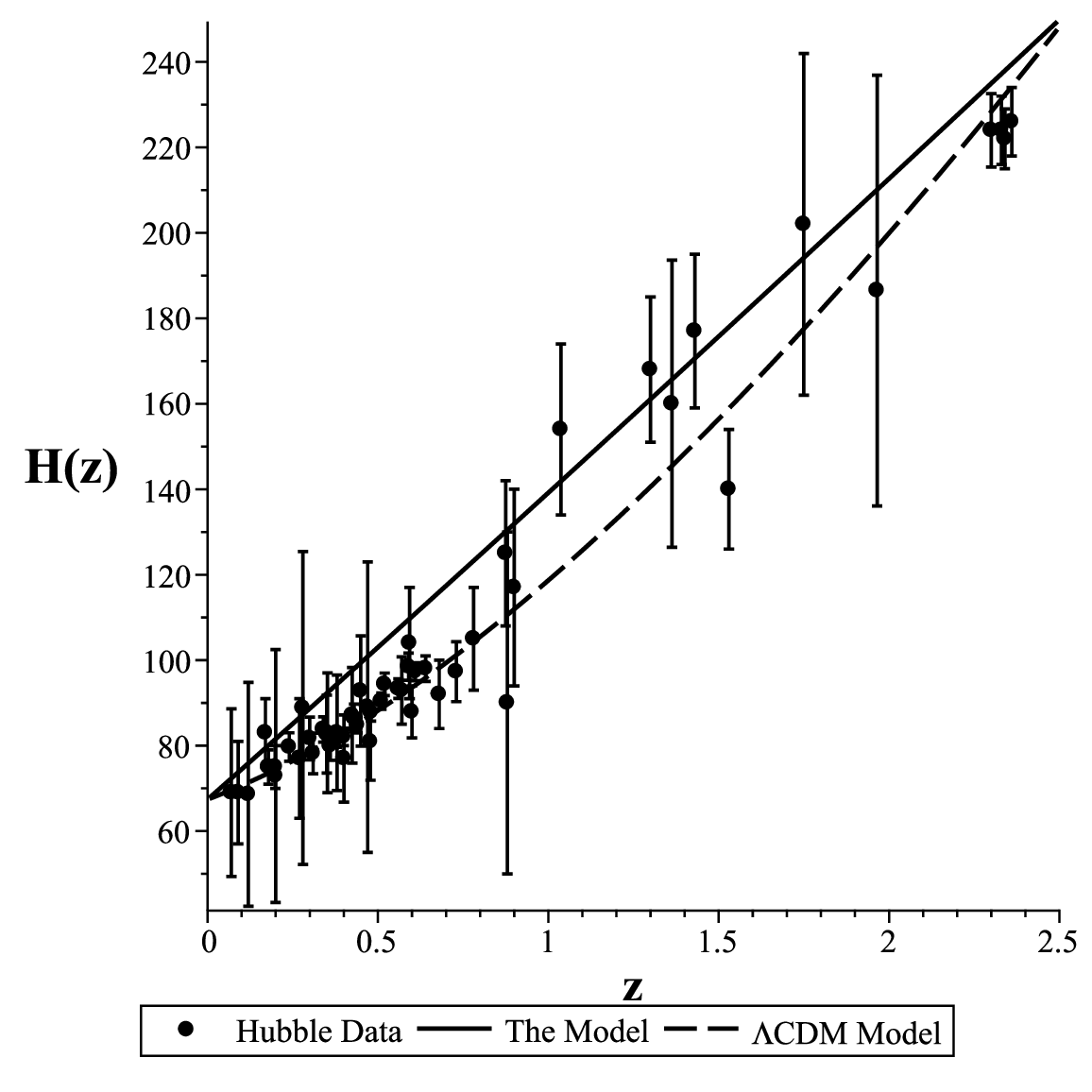}
\caption{the variations of $H(z)$ in terms of the redshift parameter.}\label{H2z}
\end{center}
\end{figure}

We now analyze the present model using observational constraints with Hubble parameter data against redshift. Observations of the expansion rate of the universe, especially Hubble parameter data $H(z)$ at different redshifts, play a fundamental role in determining and testing cosmological models. Here, relying on experimental data from galaxy observations, cluster convergence, and other independent sources, we extract statistical constraints on theoretical models using rigorous statistical tools. To examine the agreement of the present model with the experimental data, the technique of minimizing the function $\chi^2$ is used, which is defined as follows:
\begin{equation}\label{chi21}
\chi ^{2}=\sum\limits_{i=1}^N
\frac{[H_{th}(z_{i};\overrightarrow{\theta})-H_{obs}(z_{i})]^{2}}{
\sigma _{i}^{2}}.
\end{equation}
where $\overrightarrow{\theta}$ are the free parameters of the model, $H_{th}=H_0  (1+z)^{\frac{1}{m}}$ denotes the theoretical value of the Hubble parameter obtained by our model whereas $H_{obs}$ represents its observed value and $\sigma_{i}$ represents the standard deviation, and $N=53$ is the number of data points. By minimizing $\chi^2(m)$, we obtain the best-fit value of $m$, which characterizes the expansion history of the universe. This result is then used to reconstruct the torsion scalar $T = -6H^2(z)$, and subsequently the function $ f(T)$ in explicit form. The fitting confirms that the model is consistent with current Hubble data and provides a viable framework for describing late time cosmic acceleration. Next, using the 53 Hubble parameter data available in Tab. \ref{H(z)} \cite{Zhang14, Jimenez03, Simon05, Moresco12, Gaztanaga09, Oka14, Wang17, Chuang13, Alam17, Moresco16, Ratsimbazafy17, Anderson-2014, Blake12, Stern10, Moresco15, Busca13, Bautista17, Delubac15, FontRibera14}, the best-fit is obtained as $m=0.956$, from which we can immediately obtain the present age of the universe from Eq. \eqref{Eq25-2} as $t_0 = 13.87 ~Gyr$. The result of this fitting is shown in Fig. \ref{H2z} in comparison with the $\Lambda$CDM model. In Fig. \ref{H2z}, it can be seen that the 53 Hubble data points are taken from Tab. \ref{H(z)}, and the solid line graph corresponds to the current power-law model, and the dashed line graph represents the $\Lambda$CDM model.
\begin{table}[h]
\caption{The Hubble parameter data set in terms of redshift parameter and their uncertainty values, units of $H(z)$ and $\sigma_H$ are $km~ s^{-1} ~Mpc^{-1}$.} % title of Table
\centering % used for centering table
\begin{tabular}{||c | c | c | c | c || c | c | c | c | c||} % centered columns (10 columns)
\hline\hline %inserts double horizontal lines
No. & Redshift & H(z) & $\sigma_{H}$ & Ref. & No. & Redshift & H(z) & $\sigma_{H}$ & Ref.\\ [0.5ex] % inserts table
%heading
\hline % inserts single horizontal line
1. & 0.07 & 69.0 & 19.6 & \cite{Zhang14} & 28. & 0.510 & 90.4 & 1.9 &  \cite{Alam17}\\
2. & 0.09 & 69.0 & 12.0 & \cite{Jimenez03} & 29. & 0.52 & 94.35 & 2.64 &  \cite{Wang17}\\
3. & 0.12 & 68.6 & 26.2 & \cite{Zhang14} & 30. & 0.56 & 93.34 & 2.3 & \cite{Wang17}\\
4. & 0.17 & 83 & 8 &  \cite{Simon05}  & 31. & 0.57 & 92.9 & 7.855 &  \cite{Anderson-2014}\\
5. & 0.179 & 75.0 & 4.0 & \cite{Moresco12} & 32. & 0.59 & 98.48 & 3.18 &  \cite{Wang17}\\
6. & 0.199 & 75.0 & 5.0 &  \cite{Moresco12} & 33. & 0.593 & 104.0 & 13.0 & \cite{Moresco12}\\
7. & 0.200 & 72.9 & 29.6 &  \cite{Zhang14} & 34. & 0.6 & 87.9 & 6.1 &  \cite{Blake12} \\
8. & 0.24 & 79.69 & 3.32 &  \cite{Gaztanaga09} & 35. & 0.610 & 97.3 & 2.1 &  \cite{Alam17} \\
9. & 0.27 & 77 & 14 &  \cite{Simon05}  & 36. & 0.64 & 98.02 & 2.98 &  \cite{Wang17}\\
10. & 0.280 & 88.8 & 36.6 & \cite{Zhang14} & 37. & 0.680 & 92.0 & 8.0 & \cite{Moresco12}\\
11. & 0.30 & 81.7 & 5.0 &  \cite{Oka14}  & 38. & 0.73 & 97.3 & 7 &  \cite{Blake12}\\
12. & 0.31 & 78.18 & 4.74 &  \cite{Wang17} & 39. & 0.781 & 105.0 & 12 &  \cite{Moresco12}\\
13. & 0.34 & 83.8 & 2.96 &  \cite{Gaztanaga09} & 40. & 0.875 & 125 & 17 &  \cite{Moresco12}\\
14. & 0.35 & 82.7 & 9.1 &  \cite{Chuang13} & 41. & 0.880 & 90.0 & 40.0 &  \cite{Stern10}\\
15. & 0.352 & 83 & 14 &  \cite{Moresco12} & 42. & 0.900 & 117 & 23.0 &  \cite{Simon05}\\
16. & 0.36 & 79.94 & 3.38 &  \cite{Wang17} & 43. & 1.037 & 154 & 20 &  \cite{Moresco12}\\
17. & 0.38 & 81.5 & 1.9 &  \cite{Alam17} & 44. & 1.300 & 168 & 17 &  \cite{Simon05}\\
18. & 0.3802 & 83.0 & 13.5 &  \cite{Moresco16} & 45. & 1.363 & 160 & 33.6 & \cite{Moresco15}\\
19. & 0.40 & 82.04 & 2.03 &  \cite{Wang17} & 46. & 1.430 & 177 & 18 &  \cite{Simon05}\\
20. & 0.4004 & 77 & 10.2 &  \cite{Moresco16} & 47. & 1.530 & 140 & 14 &  \cite{Simon05}\\
21. & 0.4247 & 87.1 & 11.2 &  \cite{Moresco16} & 48. & 1.750 & 202 & 40 &  \cite{Simon05}\\
22. & 0.43 & 86.45 & 3.27 &  \cite{Gaztanaga09} & 49. & 1.965 & 186.5 & 50.4 & \cite{Moresco15}\\
23. & 0.44 & 84.81 & 1.83 &  \cite{Wang17} & 50. & 2.30 & 224 & 8.6 &  \cite{Busca13}\\
24. & 0.4497 & 92.8 & 12.9 &  \cite{Moresco16} & 51. & 2.33 & 224 & 8 & \cite{Bautista17}\\
25. & 0.470 & 89 & 34 & \cite{Ratsimbazafy17} & 52. & 2.340 & 222 & 7 &  \cite{Delubac15}\\
26. & 0.4783 & 80.9 & 9.0 &  \cite{Moresco16} & 53. & 2.360 & 226 & 8 &  \cite{FontRibera14}\\
27. & 0.48 & 87.79 & 2.03 &  \cite{Wang17} &  &  &  &  &  \\
[1ex] % [1ex] adds vertical space
\hline %inserts single line
\end{tabular}
\label{H(z)} % is used to refer this table in the text
\end{table}

In what follows, we consider that the source of dark energy arises from the $f(T)$ gravity and the NA. Therefore, we correspond Eq. \eqref{Eq18} with energy density of dark energy $\rho_{de}$ and obtains as
\begin{equation}\label{Eq27}
 \rho _{DE} \equiv \rho _{de}= \frac{3 {n^2} (1 - m)^2 H_0^{2}}{\kappa ^2 \, m^{2}}\, (1+z)^{\frac{2(1 - m)}{m}},
\end{equation}
where is used from relations
\begin{subequations}\label{Eq28}
\begin{eqnarray}
 &T =   -\frac{6 m^2}{t^2},\label{28-1}\\
 &\dot H =  - \frac{m}{t^2},\label{28-2}\\
 &\tau  = \int_0^t {\frac{{dt}}{a}}  = \frac{m^m}{(1-m) H_0^m}\, t^{1 - m}.\label{28-2}
\end{eqnarray}
\end{subequations}

Now, by entering equations \eqref{Eq13-11}, \eqref{Eq22}, \eqref{Eq26}, and \eqref{Eq27}in Eq. \eqref{Eq14-1}, we can obtain the form of the function $f(T(z))$ as follows:
\begin{eqnarray}\label{Eq29}
&f(z) = \frac{2 \kappa^2 \rho_{bm_0}}{1-3m(1+\omega_{bm})}(1+z)^{3(1+\omega_{bm})} + \frac{6 n^2 H_0^2 (1-m)^2}{m^2 (2m-1)} (1+z)^{\frac{2(1-m)}{m}}\notag\\
&+\frac{2 \kappa^2 b^2 \rho_{dm_0}}{\omega_{dm}}\, _2F_1\left(1, \frac{1}{3m(1-b^2)}; 1+\frac{1}{3m(1-b^2)}; \frac{(1-b^2)}{\omega_{dm}}(1+z)^{-3(1-b^2)}\right).
\end{eqnarray}

The relationship obtained for \( f(z)\) is obtained from the modified Friedmann equation in the gravitational framework \( f(T)\). This function is explicitly dependent on the redshift \( z\) and consists of three independent parts, each of which is related to one of the components of the universe:

\textbf{The first term:} This part is related to the observable baryonic matter, which is characterized by the EoS \( \omega_{bm} \) and the initial density \( \rho_{bm0} \). Its behavior corresponds to the power \( (1 + z)^3 \) in the uncompressed state (dust).

\textbf{The second term:} This part comes from the NA dark energy model, which depends on the parameters \( n \), \( m \), and \( H_0 \). Its nonlinear power represents the dynamical behavior of dark energy in the cosmological NA framework.

\textbf{The third term:} This part is related to the dark matter of the BEC type coupled with the interaction term \( Q \). The presence of the hypergeometric function \( {}_2F_1 \) indicates an analytical solution of the continuity equation with interaction and has a complex dependence on the parameters \( b \), \( \omega_{dm} \), and \( \rho_{dm_0}\).

The importance of this explicit relation for \(f(z) \) allows us to extract the function \(f(T) \) in a reconstructed form from observational data and physical models, and to analyze the dynamical behavior of the universe without assuming an arbitrary form for \(f(T) \), but rather using the physical components of the model, and finally, to compare the model with Hubble data and other observations. In that case, using Eqs. \eqref{Eq7} and \eqref{Eq26}, which leads to $1+z = \left(\frac{-T}{6 H_0^2}\right)^{\frac{m}{2}}$, we convert $f(z)$ to the function $f(T)$ as follows:
\begin{eqnarray}\label{Eq29-1}
&f(T) = \frac{2 \kappa^2 \rho_{bm_0}}{1-3m(1+\omega_{bm})}\left( \frac{-T}{6 H_0^2} \right)^{\frac{3m(1 + \omega_{bm})}{2}} + \frac{6 n^2 H_0^2 (1-m)^2}{m^2 (2m-1)} \left( \frac{-T}{6 H_0^2} \right)^{1 - m}\notag\\
&+\frac{2 \kappa^2 b^2 \rho_{dm_0}}{\omega_{dm}}\, _2F_1\left(1, \frac{1}{3m(1-b^2)}; 1+\frac{1}{3m(1-b^2)}; \frac{(1-b^2)}{\omega_{dm}}\left( \frac{-T}{6 H_0^2} \right)^{- \frac{3m(1 - b^2)}{2}}\right).
\end{eqnarray}

\begin{figure}[h]
\begin{center}
\includegraphics[scale=.4]{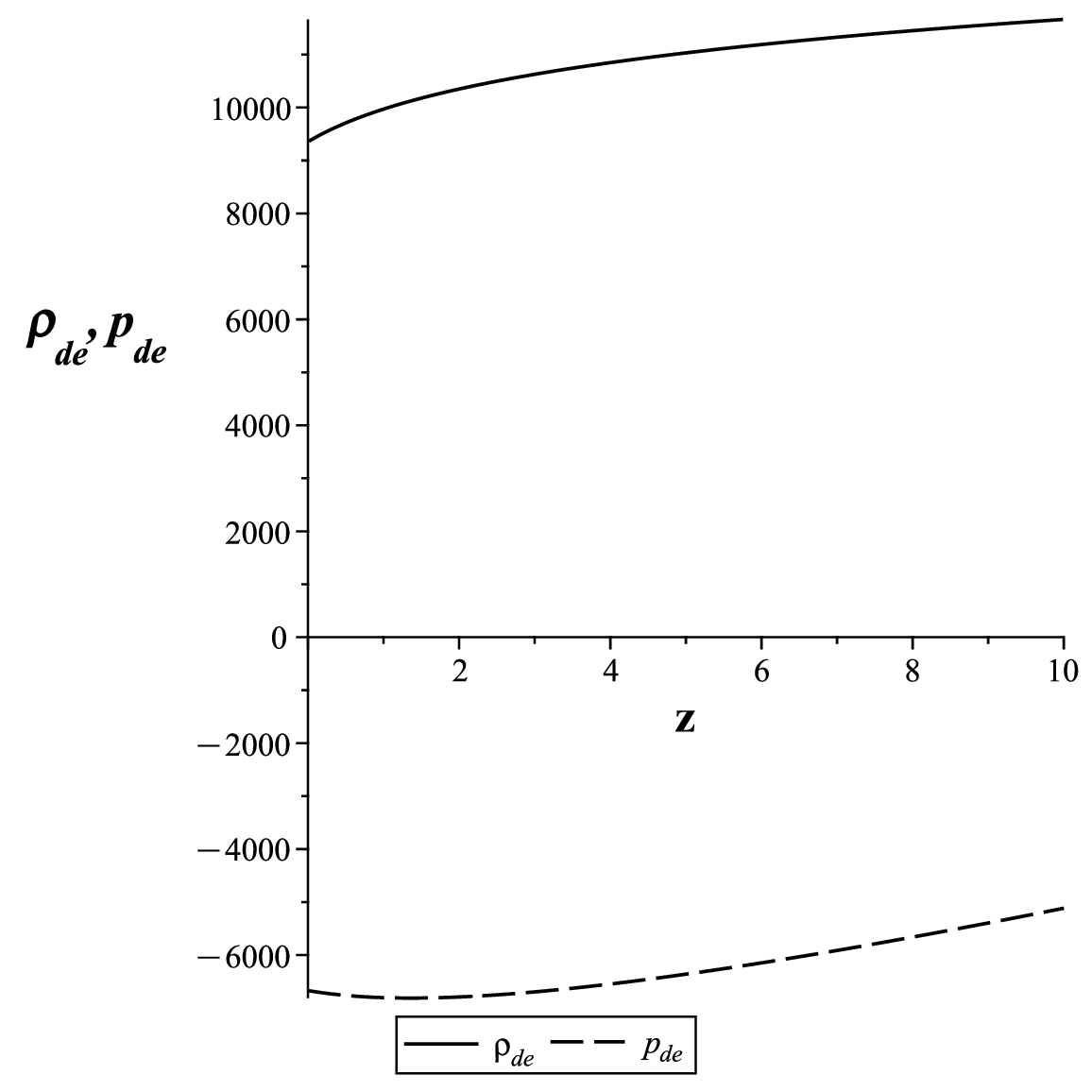}
\caption{the variations of $\rho_{de}$ and $p_{de}$ in terms of the redshift parameter.}\label{figure1}
\end{center}
\end{figure}

\begin{figure}[h]
\begin{center}
\includegraphics[scale=.4]{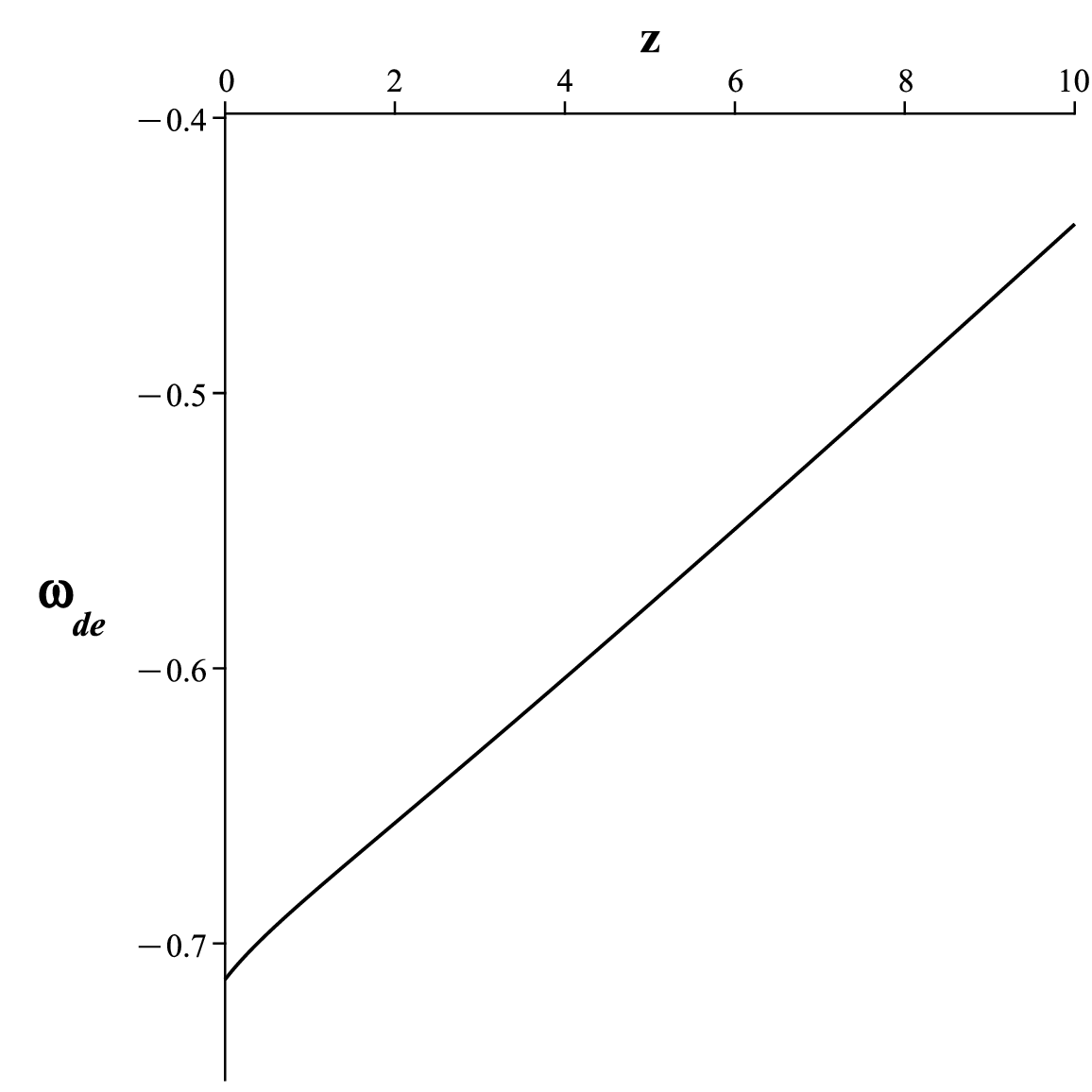}
\caption{the variations of $\omega_{de}$ in terms of the redshift parameter.}\label{figure2}
\end{center}
\end{figure}

In what follows, we can obtain the dark energy pressure $p_{de}$ by inserting Eq. \eqref{Eq27} into the continuity equation \eqref{Eq13-3}, and then we obtain the varition of $\rho_{de}$, $p_{de}$, and $\omega_{de}$ for dark energy component in terms of redshift parameter as shown in Figs. \ref{figure1} and \ref{figure2}. The choice of values of free parameters in cosmological models plays a crucial role in their physical validity and agreement with observational data. The combination of these parameters is chosen in such a way that the reconstructed function \(f(T)\) is not only consistent with experimental data, but also has theoretically acceptable behavior at different epochs of the universe, including the transition from the matter era to the accelerated era. Therefore, the free parameters are selected as $n = 18$, $b = 0.807$, $\omega_{dm} = 2.5$, $\rho_{dm_0} = 4873$, $\omega_{bm} = 1$, and $\rho_{bm_0} = 545$ which $n$, $b$, $\omega_{dm}$, and $\omega_{bm}$ are dimensionless quantities, and $\rho_{dm_0}$ and $\rho_{bm_0}$ are dimension $M^{4}$ in Planck units. We can see in Figs. \ref{figure1} that the value of energy density decrease from a high positive value in the early universe to a lower positive value in the late universe ($z = 0$), also the value of pressure decreases from a high positive value in the early universe to a lower negative value in the late universe ($z = 0$).

We note that the EoS parameter is a very important parameter so that can understand at a glance how the dynamics of the universe are evolving from the early time to the late time. Therefore, we notice that the matter period and the accelerated pase period for the universe can be expressed as $0 < \omega < \frac{1}{3}$ and $-1 < \omega < -\frac{1}{3}$, respectively. Hence, Fig. \ref{figure2} shows us that the universe evolves from the matter-dominated era to the accelerated expansion era in which the value of EoS is about $-0.71$ in the present time ($z=0$), indicating that the universe enters a thermodynamically unstable phase in the late-time evolution. 

\begin{figure}[h]
\begin{center}
\includegraphics[scale=.35]{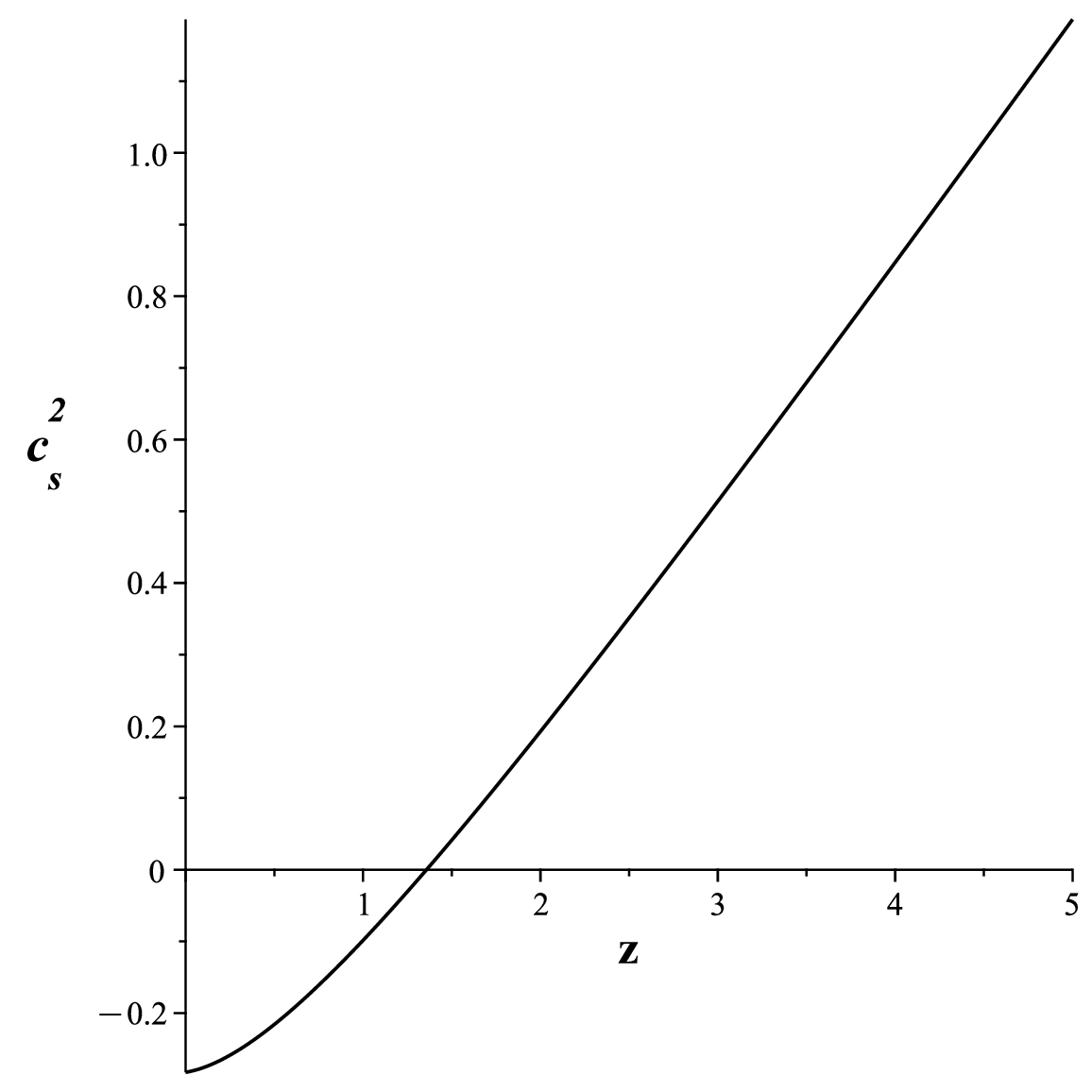}
\caption{The graph of the sound speed (dimensional quantity) in terms of redshift parameter with the same values of $m = 0.956$, $n = 18$, $b = 0.807$, $\omega_{dm} = 2.5$, $\rho_{dm_0} = 4873$, $\omega_{bm} = 1$, and $\rho_{bm_0} = 545$.}\label{figure3}
\end{center}
\end{figure}

In what follows, we explore the stability of the model by using the sound speed parameter which is derived from thermodynamic considerations. For this reason, we consider the universe to be an adiabatic thermodynamic system where no heat or mass is transferred out, meaning that the perturbation has zero entropy. Therefore, the sound speed parameter $c_s^2$ is defined in the following form
\begin{equation}\label{Eq30}
c_s^2=\frac{\partial p_{de}}{\partial \rho_{de}} = \frac{\partial_z p_{de}}{\partial_z \rho_{de}},
\end{equation}
where the index $z$ indicates derivative with respect to redshift parameter. We plotted the sound speed parameter as a function of redshift parameter as shown in Fig. \ref{figure3}. We note that the conditions of larger and smaller sound speed parameters, i.e., $c_s^2 > 0$ and $c_s^2 < 0$, indicate stability and instability, respectively. In that case, Fig. \ref{figure3} displays that the universe begins from a stability phase and then at the threshold of the accelerated phase enters a non-stability phase in late time. Therefore, this variation represents that the energy density of dark energy in the late universe is not in a controlled growth.

At the end, we test the present model by using the density parameters of dark energy, dark matter, and baryonic matter. For this purpose, we account the corresponding density parameters as follows:
\begin{subequations}\label{Eq31}
\begin{eqnarray}
 &\Omega_{bm_0} = \frac{\rho_{bm_0}}{\rho_c},\label{Eq31-1}\\
 &\Omega_{dm_0} = \frac{\rho_{dm_0}}{\rho_c},\label{Eq31-2}\\
 &\Omega_{de_0} = \frac{\rho_{de_0}}{\rho_c},\label{Eq31-2}
\end{eqnarray}
\end{subequations}
where $\rho_c = 3 H^2 / \kappa^2$ is the critical density, and index zero is introduced the present values of the corresponding quantity. Now, by substituting the relationships related to energy density \eqref{Eq13-11}, \eqref{Eq22} and \eqref{Eq27} in the above relationship, the contribution of the universe components in the late universe is calculated according to table \ref{table1}. The contents of Table \ref{table1} are calculated for the late time state ($z=0$), and show that the matter-energy contribution into the universe is consistent with the observational data \cite{Aghanim-2017}. The energy budget values from this reference are derived from precise measurements of the anisotropies of the cosmic microwave background (CMB) and other cosmological observations.

\begin{table}[h]
\caption{The present values of the density parameters of $\Omega_{bm_0}$, $\Omega_{dm_0}$, and $\Omega_{de_0}$.} % title of Table
\centering % used for centering table
\begin{tabular}{||c |c ||} % centered columns (4 columns)
\hline\hline %inserts double horizontal lines
~$\rm{the~ density~ parameter}$~ & $\rm{the~ present~ value}$ \\ [0.5ex] % inserts table
%heading
\hline\hline % inserts single horizontal line
$\Omega_{bm_0}$ &  $0.04$ \\
 \hline% inserting body of the table
$\Omega_{dm_0}$ &  $0.27$  \\
 [1ex] % [1ex] adds vertical space
\hline %inserts single line
$\Omega_{de_0}$ &  $0.69$ \\
 [1ex] % [1ex] adds vertical space
\hline\hline %inserts single line
\end{tabular}
\label{table1} % is used to refer this table in the text
\end{table}

%####################################################################
%####################################################################
\section{Conclusion}\label{VI}

In this paper, we studied the cosmic evolution of the universe in the \( f(T) \) modified gravity framework with a smooth Friedmann-Robertson-Walker (FRW) background. We considered the universe to consist of three components: baryonic matter, dark matter, and dark energy, where dark matter was replaced by the BEC model and dark energy by the NA dark energy model. After deriving the Einstein field equations and Friedmann equations in the \( f(T) \) background, we wrote the continuity equations for each component of the universe separately and considered an interaction between dark matter and dark energy in which energy is transferred from dark matter to dark energy.

For the dynamical analysis, we chose power-law cosmology for the scale factor and reconstructed the Hubble parameter in terms of the redshift parameter \( z \). Using the 53-point Hubble parameter data, the optimal value of the power parameter \( m = 0.956 \) was obtained, which led to the calculation of the current age of the universe as \( t_0 = 13.87 ~Gyr\). This value is fully consistent with recent observational data.

Next, using the BEC model for dark matter and the NA dark energy model for dark energy, we obtained the function \( f(T) \) explicitly and reconstructed in terms of the torsion scalar \( T \). This function consists of three independent parts:\\
- The first part: related to baryonic matter with a strong dependence on \( (1 + z)^3 \),\\
- The second part: related to the dark energy of the new sensor with a nonlinear dependence on the parameters \( n \), \( m \) and \( H_0 \),\\
- The third part: related to BEC dark matter type with the presence of the interaction term and the hypergeometric function \(_2F_1 \) which represents the analytical solution of the continuity equation with interaction.

Plots of energy density \( \rho_{\text{de}} \), pressure \( p_{\text{de}} \), and EoS parameter \( w_{\text{de}} \) versus redshift showed that the universe moves from a matter-centered phase to an accelerated expansion, and the value \( w_{\text{de}} \approx -0.71 \) was obtained at the present time ($z = 0$). This value indicates that the universe enters an accelerated and unstable phase at later times.

To investigate thermodynamic stability, we calculated the speed of sound parameter \( c_s^2 \). The corresponding plot showed that the universe transitions from a stable phase at early times to an unstable phase at later times, which is consistent with the behavior of dark energy in dynamical models.

Finally, by calculating the current density parameters for baryonic matter, dark matter, and dark energy as \(\Omega_{bm_0} = 0.04 \), \( \Omega_{dm_0} = 0.27 \), and \( \Omega_{de_0} = 0.69 \), we show that the proposed model is in good agreement with recent observational data, including CMB results and large-scale structure.

It is worth noting that in the present model, the roles of these three components are physically distinct, although they are connected through the Friedmann background. The function $f(T)$ determines the geometric evolution of the universe through torsion and provides the gravitational framework. The BEC dark matter, through its nonlinear equation of state, imposes quantum corrections that mainly affect the matter-dominated period and structure formation. The NA dark energy component, in turn, drives the finite-time acceleration through the isotropic time dependence of its energy density. Hence, while these three components interact dynamically, their individual actions remain separate and contribute differently to cosmic evolution.

In addition to the numerical and analytical results obtained, it can be emphasized that the main objectives of the research have been fully realized. The proposed model was able to provide a coherent picture of the evolution of the universe by combining the three theoretical frameworks of $f(T)$ gravity, BEC dark matter, and NA dark energy, which is consistent with observational data. The reconstruction of the $f(T)$ function based on the physical components of the model and the $H(z)$ data showed that the proposed model is able to explain the cosmic acceleration without the need for a cosmological constant. Also, the analysis of thermodynamic stability and the behavior of the equation of state parameter confirmed that the transition from the matter-dominated phase to the acceleration-dominated phase is consistent with theoretical predictions. Therefore, the results of this research indicate that the assumptions and objectives set at the beginning of the work, including the geometric-quantum description of the universe and compatibility with observations, have been achieved quantitatively and qualitatively. In the future, this model can be extended by analyzing perturbations, structure formation, and CMB and BAO data to further examine its validity.

%###################################################################################################

\end{document}